\documentclass[aps,prl,reprint,groupedaddress] {revtex4-1}
\usepackage{graphicx} 
\usepackage{dcolumn} 
\usepackage{bm}
\usepackage{hyperref}
\usepackage{amsmath}

\begin{document}

\title{ Probing the neutron-skin thickness via the double-charge exchange reactions in pion-nucleus collisions }
\author{Ban Zhang$^{1}$ }
\author{Zhao-Qing Feng$^{1,2}$ }
\email{Corresponding author: fengzhq@scut.edu.cn}
\affiliation{ $^{1}$School of Physics and Optoelectronics, South China University of Technology, Guangzhou 510640, China  \\
$^{2}$State Key Laboratory of Heavy Ion Science and Technology, Institute of Modern Physics, Chinese Academy of Sciences, Lanzhou 730000, China }

\begin{abstract}
Within the framework of Lanzhou quantum molecular dynamics (LQMD) transport model, we investigate the influence of the neutron-skin thickness of neutron-rich nuclei on the pion emission in the double-charge exchange reactions near the $\triangle$-resonance energy, in particular, the rapidity and transverse momentum spectra, charged pion ratios et al. The reactions induced by pions are associated with the multiple processes of pions, nucleons and $\triangle$-resonances, correlated with the neutron-skin thickness. It is found that the kinetic energy spectra of $\pi^{-}$/$\pi^{+}$ ratio are sensitive to the neutron-skin thickness of neutron-rich nuclei for extracting the subsaturation-density symmetry energy. The double-charge exchange reactions are influenced by the neutron-skin thickness, in-medium properties of resonances, pion-nucleon potential and pion-nucleon scattering.

\end{abstract}

\maketitle

\section{I. Introduction}
In the past several decades, the dynamics of pion-induced nuclear reactions has been extensively investigated both in theories and in experiments. The investigation of pion-nucleus collisions also deepens the understanding of nuclear dynamics in heavy-ion collisions, hadron-induced nuclear reactions, high-intensity muon source etc. Pion-induced reactions offer significant advantages in comparison with the proton-induced reactions, in particular in the low-energy region, in which the inelastic pion-nucleus cross section is considerably larger than that of proton-nucleus reactions. A number of experiments on the pion-nucleus collisions concentrated on the elastic scattering, absorption, spallation reactions, charge-exchange reactions, strangeness production, hypernuclear physics et al. \cite{By52,Ca71,Da72,Wi73,Cl74,Ca76,As81,Wo92,Kr99,Cr04}. The extraction of symmetry energy from heavy-ion collisions in high-density region, particularly the pion probes near the threshold energy, have been attracted attention, i.e., the charged pion ratio, flow difference, double ratio in the isotropic reactions \cite{Fe06,Di10,Xiao09,Feng10}. Moreover, the in-medium effects of pions in heavy-ion collisions were extensively investigated \cite{Fe05,So15}, such as the cross sections of production and reabsorption, pion-nucleon potential et al. The pion optical potential influences the yields and pion distribution in phase space \cite{Sa87,En94,Ho14,Feng10a}.  The pion-induced nuclear reaction provides a new way for investigating the equation of state (EoS) from the neutron-skin thickness of neutron-rich nuclei.

In the neutron-rich nuclei, a part of neutrons tend to concentrate on the nuclear surface, forming a neutron skin, which is typically characterized by the difference in the root-mean-square (rms) radii of the neutron and proton density distributions, namely, $\Delta r_{np} = \left\langle r_n^2 \right\rangle^{\frac{1}{2}} - \left\langle r_p^2 \right\rangle^{\frac{1}{2}}$. The distribution of protons over a large range of nuclei has been well studied by measuring the electron-nucleus elastic scattering and muonic atom X-rays \cite{Fr92,Fr95,Vr87,Wo80,Wo81}. The neutron distribution inside a atomic nucleus is enhanced by the fact that the number of neutrons, N, is larger than the number of protons, Z, for both medium and heavy nuclei. The neutron-rich nuclei with large isospin asymmetry provide a connection bridge between the EoS of finite nuclei and infinite nuclear matter of neutron stars. The accurate determination of the neutron-skin thickness is related to several issues, i.e., the nuclear structure, neutron-star matter properties, atomic parity violation (PV), heavy-ion collision dynamics \cite{Ho01,Ho06,Pi07}. In particular, the recent measurement of neutron-skin thickness of $^{208}\rm{Pb}$ from weak violation in the electron-nucleus scattering \cite{Ad21} has attracted attention and enabled more discussion in nuclear structure calculations \cite{Hu22,Es21,Re22}.

Double charge exchange (DCX) reactions in the pion-induced reactions pave the way for elucidating the multiple pion-nucleon collisions on the nuclear surface. The method provides the possibility for extracting the neutron-skin thickness because of the difference of scattering cross sections. Usually, the concept of two-step sequential single-charge exchange processes has been able to explain the characteristics of DCX at low energies \cite{Be70,Gi77}. In Ref. \cite{Vi89}, the DCX reactions involving the pion induced on $^{16}\rm{O}$ and $^{40}\rm{Ca}$ were investigated and consistent with the experimental data. More data on the charged pion induced reactions on $^{16}\rm{O}$, $^{40}\rm{Ca}$, $^{103}\rm{Rh}$ and $^{208}\rm{Pb}$ were measured at the Clinton P. Anderson Meson Physics Facility (LAMPF) \cite{Wo92}. The attempt of DCX on the neutron-rich nuclei for extracting the neutron-skin thickness is promising in the future experiments. Calculations by the Giessen BUU (GiBUU) transport model manifested that the 35\% enhancement of DCX cross sections with inclusion of the neutron-skin profile for the reaction $ \pi^+ + \text{Pb} \rightarrow \pi^- +  \text{X} $ \cite{Bu06}. The neutron-skin thickness of neutron-rich nuclei might be investigated with the high-intensity pion beams at High
Intensity heavy-ion Accelerator Facility (HIAF) \cite{Ya13} and with the muon source at Initiative Accelerator Driven Subcritical System (CiADS) in Huizhou \cite{Ca24}.

In this work, we employ the Lanzhou quantum molecular dynamics (LQMD) transport model to explore the pion-nucleus collisions in the $\Delta$ resonance energies and investigate the impact of neutron-skin thickness on the pion production. Section \uppercase\expandafter{\romannumeral2} is a brief introduction of theoretical method. The observables for extracting the neutron-skin thickness are discussed in section \uppercase\expandafter{\romannumeral3}. The perspective on the measurements of the neutron-skin thickness via the low-energy pion induced reactions is summarized in \uppercase\expandafter{\romannumeral4}.

\section{II. Model description}
The LQMD transport model has been widely used in heavy-ion collisions and hadron (proton, antiproton, meson and hyperon) induced reactions \cite{Feng11,Feng12,Liu23}. In the LQMD model, the elastic and inelastic hadron-hadron collisions, the production, decay and reabsorption of resonances, as well as the interaction potential between hadrons and nucleons are self-consistently implemented. The dynamics of the nuclear fragmentation and the charge exchange reactions in pion-nucleus collisions near the $\Delta$ resonance energies has been investigated within the LQMD \cite{Feng16}. The neutron-skin thickness of neutron-rich nuclei in the pion-induced reactions is to be investigated for the first time.

\begin{figure*}[t]
\centering
\includegraphics[width=\linewidth]{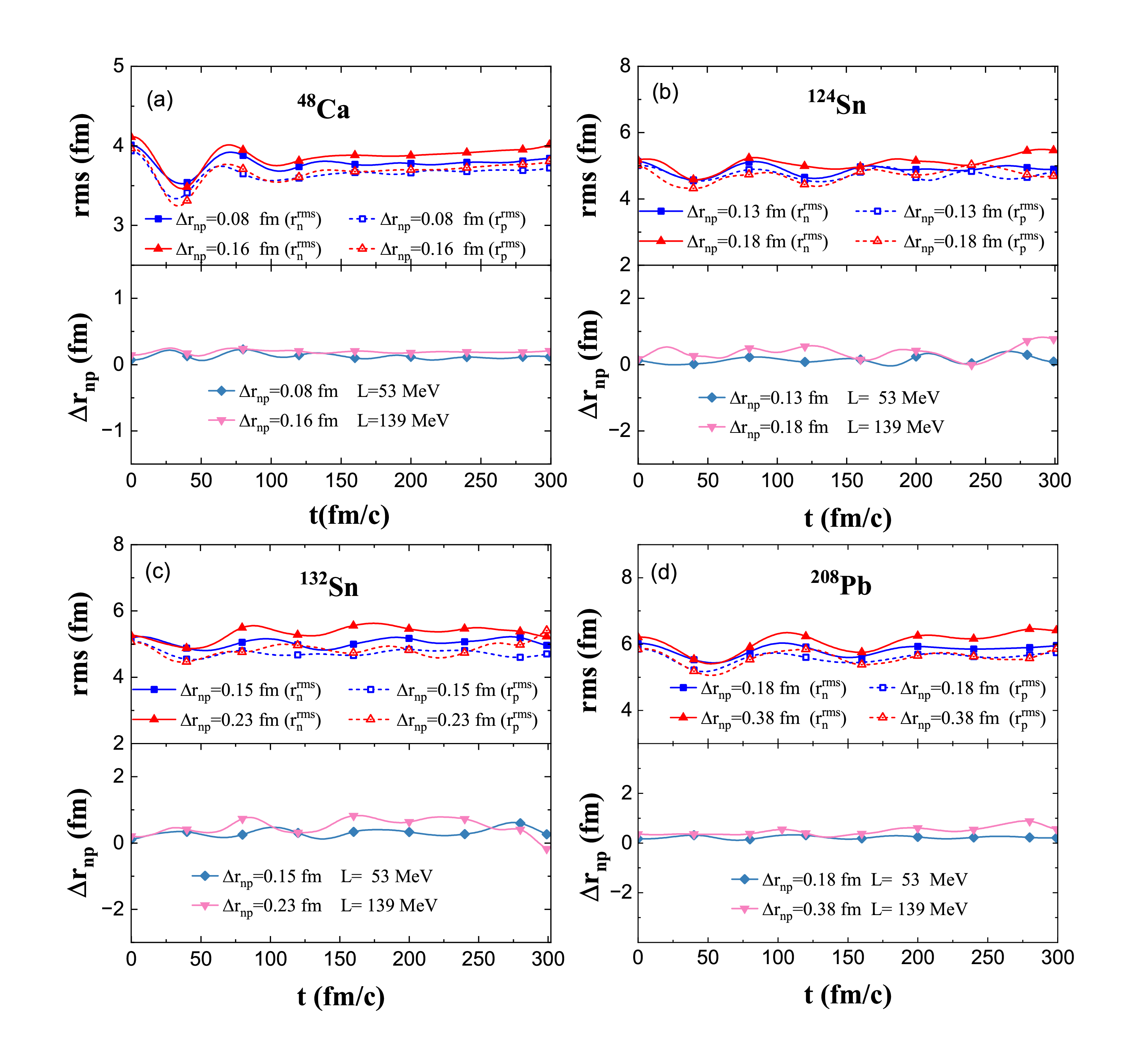}
\caption{\label{1} The time evolution of rms radii and neutron-skin thicknesses with the different slope parameter of symmetry energy for $^{48}\rm{Ca}$, $^{124}\rm{Sn}$, $^{132}\rm{Sn}$ and $^{208}\rm{Pb}$. }
\end{figure*}

\subsection{A. Interaction Hamiltonian}
In the LQMD model, each nucleon of reaction system is represented by a Gaussian wave-packet as
\begin{equation}
\phi_i(\textbf{r},t) = \frac{1}{(2\pi L)^\frac{3}{4}} \exp\left[ -\frac{(\textbf{r}-\textbf{r}_i(t))^2}{4L} + \frac{i\textbf{p}_i(t)\cdot\textbf{r}}{\hbar}\right],
\end{equation}
where the $L$ represents the square of the wave-packet width, which is relative to the mass number of projectile or target nucleus as $\sqrt{L}=0.08A^{\frac{1}{3}}+0.99$ fm. The variables $\textbf{r}_i(t) $ and $\textbf{p}_i(t) $ denote the center position of the $i-$th wave-packet in the coordinate and momentum space, respectively. The total wave-function of the reaction system is the direct product of each nucleon wave-function by
\begin{equation}
\Phi(\textbf{r},t)=\prod_{i}\phi_i(\textbf{r},\textbf{r}_i,\textbf{p}_i, t).
\vspace{-0.5em}
\end{equation}
The temporal evolution of all particles is governed by self-consistently generated mean-field potentials, the coordinates and momentum of the $i-$th particle evolve according to a Hamiltons equations of motion, which are given by
\begin{equation}
\dot{\bm{r}_i}=\frac{\partial H }{\partial\bm{p}_i},    \qquad    \dot{\bm{p}_i}=-\frac{\partial H }{\partial\bm{r}_i}.
\end{equation}
The Hamiltonian consists of the relativistic energy, Coulomb interaction, and local interaction as follows
\begin{equation}
H_B = \sum_i \sqrt{\mathbf{p}_i^2 + m_i^2} + U_{Coul} + U_{loc},
\end{equation}
here the $\mathbf{p}_i$ and $m_i$ represent the momentum and the mass of the baryons.

The local interaction potential is evaluated by the energy-density functional as
\begin{equation}
U_{loc}=\int V_{loc}[\rho(\bm{r})]d\bm{r},
\end{equation}
with
\begin{eqnarray}
 V_{loc}& = \frac{\alpha}{2}\frac{\rho^2}{\rho_0} + \frac{\beta}{1+\gamma}\frac{\rho^{1+\gamma}}{\rho_0^\gamma} + \frac{1}{2} C_{sym} \left(\frac{\rho}{\rho_0}\right)^{\gamma_s} \rho\delta^2  \nonumber\\
\vspace{+0.5em}
 & +\frac{g_{sur}}{2\rho_0}(\nabla \rho)^2+\frac{g_{sur}^{iso}}{2\rho_0}[\nabla (\rho_n-\rho_p)]^2,
\end{eqnarray}
where $\delta=(\rho_n- \rho_p)/(\rho_n+ \rho_p)$ denotes the isospin asymmetry, and $ \rho_n$ and $\rho_p$ denote the neutron and proton densities, respectively. In the LQMD model, the parameters are defined as follows: $\alpha = -226.5\ \rm{MeV}$ , $\beta = 173.7\ \rm{MeV}$  and $\gamma = 1.039$.  These correspond to a compression modulus of $ K = 230\ \rm{MeV}$  for isospin symmetry nuclear matter at the saturation density ($\rho_0 = 0.16\ fm^{-3}$). Additionally, the surface symmetry energy parameters are taken as $g_{sur}=23 \ \rm{MeV} \ fm^2$ and $g_{sur}^{iso}=-2.7  \ \rm{MeV} \ fm^2$ for describing the binding energies of finite nuclei. The symmetry energy is composed of the kinetic energy from the nucleonic Fermi motion and the local density dependent term as:
\begin{equation}
E_{sym}(\rho) = \frac{1}{3} \frac{\hbar^2}{2m} \left(\frac{3}{2} \pi^2 \rho \right)^{2/3} + \frac{1}{2} C_{sym} \left(\frac{\rho}{\rho_0}\right)^{\gamma_s},
\end{equation}
in which the parameter $C_{sym}$ is taken to be 38 MeV and $\gamma_s$ can be taken to be 0.5, 1, and 2, representing the soft symmetry energy, the linear symmetry energy, and the hard symmetry energy, corresponding to the slope parameters $[L(\rho _{0}) = 3\rho_{0} dE_{sym}(\rho)/d\rho|_{\rho=\rho_{0}}]$ of 53, 82, and 139 MeV, respectively.

The pion dynamics is influenced by the mean-field potential in nuclear medium. Analogously to baryons, the Hamiltonian of pions is represented as
\begin{equation}
H_M = \sum_{i=1}^{N_M} \left[ V_i^{Coul} + \omega(p_i, \rho_i) \right],
\end{equation}
where $V_i^{Coul}$ is Coulomb interaction potential energy:
\begin{equation}
V_i^{Coul} = \sum_{j=1}^{N_B} \frac{e_i e_j}{r_{ij}},
\end{equation}
where ${e_i}$ ($ {e_j}$) denotes the charge number, and the relative distance between the charged particles is represented as ${r_{ij}}$, $N_M$ and $N_B$  represent the are the total number of pions and baryons including charged resonances, $\omega(p_i, \rho_i)$  represents the energy of a meson within the nuclear medium, the contributions to it from the isoscalar and isovector are as follows \cite{Feng1517}:
\begin{equation}
\omega_{\pi}\left(p_{i},\rho_{i}\right)=\omega_{\text{isoscalar}}\left(p_{i},\rho_{i}\right)+C_{\pi}\tau_{z}\delta\left(\frac{\rho}{\rho_{0}}\right)^{\gamma_{\pi}}
\end{equation}
The coefficient $C_{\pi} = \frac{\rho_{0} \hbar^{3}}{4 f_{\pi}^{2}} = 36 \text{ MeV}$ is taken from fitting the experimental data of pion-nucleus scattering \cite{Fr07}. The isospin quantities are taken to be $\tau_z = $ 1, 0,and -1 for $\pi^-$, $\pi^0$, and $\pi^+$, respectively. The isospin asymmetry $\delta = \frac{(\rho_n - \rho_p)}{(\rho_n + \rho_p)}$ and the quantity $\gamma_{\pi}$ adjusts the stiffness of isospin splitting of the pion-nucleon potential. The isospin stiffness of $\gamma_{\pi}= 2 $ is taken in the work, which manifests the difference of $\pi^-$-neutron and $\pi^+$-neutron scattering in the dense neutron-rich matter. The isoscalar part of the pion self-energy in the nuclear medium is evaluated via the $\Delta$-hole model.

\subsection{B. Correlation of initialization of target nucleus and symmetry energy}
For the neutron-rich nuclei, the density distributions of neutrons and protons exhibit significant variation. Within the LQMD model, distinguishing between neutrons and protons is crucial for accurately sample stable nucleus with a neutron-skin. In the initialization, the neutron and proton density profiles of target nuclei are calculated by the well-known Skyrme Hartree-Fock method. Consequently, the density distribution is sampled with a two-parameter Fermi form as
\begin{equation}
\rho^{T}_i=\frac{\rho^{T}_{0i}}{1+\exp(\frac{r-R^{T}_i}{a_i})}, \ i=n,p\ ,
\end{equation}
where $\rho^{T}_{0n(p)}$ is the central density of neutrons (protons) in the nucleus, $R^{T}_{n(p)}$, $a_{n(p)}$ are the radius and diffusion coefficient of neutrons (protons) density distributions, respectively. In the system, it is taken into account that the number of particles is conserved, $N=\int \rho^{T}_n(\bm{r})d^3r$ and $Z=\int \rho^{T}_p(\bm{r})d^3r$. We maintain a constant proton density distribution while varying diffusion coefficient of the neutron density distribution, an increased diffusion coefficient results in a larger rms radius for neutron, thereby enhancing the neutron-skin thickness. For $^{48}\rm{Ca}$, $^{124}\rm{Sn}$, $^{132}\rm{Sn}$ and $^{208}\rm{Pb}$, both the proton and neutron density distributions are parameterized with the parameters given in Table \ref{tab:1}.
\begin{table}[ht]
\caption{Parameters of the Fermi distribution for $^{48}\rm{Ca}$, $^{124}\rm{Sn}$, $^{132}\rm{Sn}$ and $^{208}\rm{Pb}$, respectively.}
\label{tab:1}
\renewcommand{\arraystretch}{1.5}
\begin{tabular}{cccccccc}
\hline
\hline
 & \(p_{0p} (\text{fm}^{-3})\) &\(p_{0n} (\text{fm}^{-3})\)& \(R_p (\text{fm})\) & \(R_n (\text{fm})\) & $a_p (\text{fm})$ & $a_n (\text{fm})$ & $L (\text{MeV})$ \\[0.5ex]
\hline
$^{48}\rm{Ca}$ & 0.0731 & 0.0872 & 3.790 & 4.062 & 0.54 & 0.49 & 53\\[0.5ex]
                             & 0.0731 & 0.0854 & 3.790 & 4.062 & 0.54 & 0.53 & 139\\[0.5ex]
$^{124}\rm{Sn}$ & 0.0648 & 0.0964 & 5.576 & 5.492 & 0.44 & 0.57& 53\\[0.5ex]
                             & 0.0648 & 0.0951 & 5.576 & 5.502 & 0.44 & 0.59 & 139\\[0.5ex]
$^{132}\rm{Sn}$ & 0.0620 & 0.0994 & 5.670 & 5.650 & 0.43 & 0.54 & 53\\[0.5ex]
                              & 0.0620 & 0.0957 & 5.670 & 5.710 & 0.43 & 0.57 & 139\\[0.5ex]
$^{208}\rm{Pb}$ & 0.0633 & 0.090 & 6.638 & 6.780 & 0.506 & 0.57 & 53\\[0.5ex]
                               & 0.0633 & 0.090 & 6.638 & 6.730 & 0.506 & 0.66 & 139\\[0.5ex]
\hline
\hline
\end{tabular}
\end{table}

The space coordinates of nucleons are derived from a uniform sampling of the nuclear density distribution, ensuring the discrepancy between the sampled rms radius and the initial rms radius is less than 0.1 fm, resulting in a uniform nucleon distribution. The Fermi momentum is calculated using the formula $P_F^i(\bm{r})= \hbar [3\pi^2\rho^{T}_i(\bm{r})]^\frac{1}{3}$, $i=n,p$, and the momentum coordinates are derived from random sampling within the range [0, $P_F^i$]. Subsequent sampling is performed until the difference between the nuclear binding energy and its experimentally determined value falls within an acceptable range. The stability of the initial nucleus is assessed and confirmed to be sustainable up to 300 fm/c under mean-field evolution. Fluctuations are observed in the rms radii and neutron-skin thicknesses derived from this methodology. However, these variations remain within the desired limits. Fig.\ref{1} presents the rms radii of neutrons and protons, as well as the neutron skin, for $^{48}\rm{Ca}$, $^{124}\rm{Sn}$, $^{132}\rm{Sn}$ and $^{208}\rm{Pb}$, respectively. It is evident that the rms radius of protons remains essentially constant across various neutron skin thickness settings. Even under the minor fluctuations, a distinct neutron-skin thickness is sustained up to a time evolution of 300 fm/c.

\subsection{C. Reaction channels induced by pions}
The probability of two-particle collisions in a channel is determined using a Monte Carlo procedure based on the relative distance smaller than the scattering radius. The channels associated with the pion-induced reactions are as follows:
\begin{equation}
\begin{array}{l}
N\pi \leftrightarrow \Delta, \quad N\pi \leftrightarrow N^{*}, \quad NN\pi(s-\text{state}) \leftrightarrow NN, \\
N\Delta \leftrightarrow NN, \quad NN^{*} \leftrightarrow NN, \quad \Delta\Delta \leftrightarrow NN.
\end{array}
\end{equation}
The momentum-dependent decay widths are used for the resonances of $\Delta$	(1232) and $N^{*}$(1440) \cite{Hu94}. We have taken a constant width of $\Gamma = 150 \text{ MeV}$ for the $N^*$(1535) decay. The cross section of pion-nucleon scattering is evaluated with the Breit-Wigner formula as follows
\begin{equation}
\sigma_{\pi N \rightarrow R}(\sqrt{s}) = \sigma_{\text{max}} |\mathbf{p}_0/\mathbf{p}|^2 \frac{0.25 \Gamma^2(\mathbf{p})} {0.25 \Gamma^2(\mathbf{p}) + (\sqrt{s} - m_0)^2}
\end{equation}
where $\mathbf{p}$ and $\mathbf{p_0}$ are the momenta of pions at the center-of-mass energies of $\sqrt{s}$ and $m_0$, respectively. The $m_0$ is the resonance mass, e.g., 1.232, 1.44, and 1.535 GeV for $\Delta$(1232), $N^{*}$(1440), and $N^*$(1535), respectively. The $\Delta$(1232) resonance dominates the DCX reactions at the energy considered in this work.
The maximum cross section $\sigma_{max}$ is taken by fitting the total cross sections of the available experimental data in pion-nucleon collisions with the Breit-Wigner form of resonance formation as \cite{Feng16}
\begin{align}
&\sigma_{\text{max}} (\pi^+ + p \rightarrow \Delta^{++}) = \sigma_{\text{max}} (\pi^- + n \rightarrow \Delta^{-}) = 200 \text{ mb}, \\
&\sigma_{\text{max}} (\pi^0 + p \rightarrow \Delta^{+}) = \sigma_{\text{max}} (\pi^0 + n \rightarrow \Delta^{0}) = 133.3 \text{ mb}, \\
&\sigma_{\text{max}} (\pi^- + p \rightarrow \Delta^{0}) = \sigma_{\text{max}} (\pi^+ + n \rightarrow \Delta^{+}) = 66.7 \text{ mb}
\end{align}
and
\begin{align}
\sigma_{\text{max}} (\pi^{0}+p\rightarrow N^{*+}(1440)) &= \sigma_{\text{max}} (\pi^{0}+n\rightarrow N^{*0}(1440))  \\
& = 12 \text{mb}, \\
\sigma_{\text{max}} (\pi^{-}+p\rightarrow N^{*0}(1440))  &= \sigma_{\text{max}} (\pi^{+}+n\rightarrow N^{*+}(1440))   \\
& = 24 \text{mb}, \\
\sigma_{\text{max}} (\pi^{0}+p\rightarrow N^{*+}(1535)) &= \sigma_{\text{max}} (\pi^{0}+n\rightarrow N^{*0}(1535))   \\
& = 16 \text{mb}, \\
\sigma_{\text{max}} (\pi^{-}+p\rightarrow N^{*0}(1535))  &= \sigma_{\text{max}} (\pi^{+}+n\rightarrow N^{*+}(1535))   \\
& = 32 \text{mb}.
\end{align}

\section{III. Results and discussion}

\begin{figure}[ht]
\centering
\includegraphics[width=\linewidth,keepaspectratio]{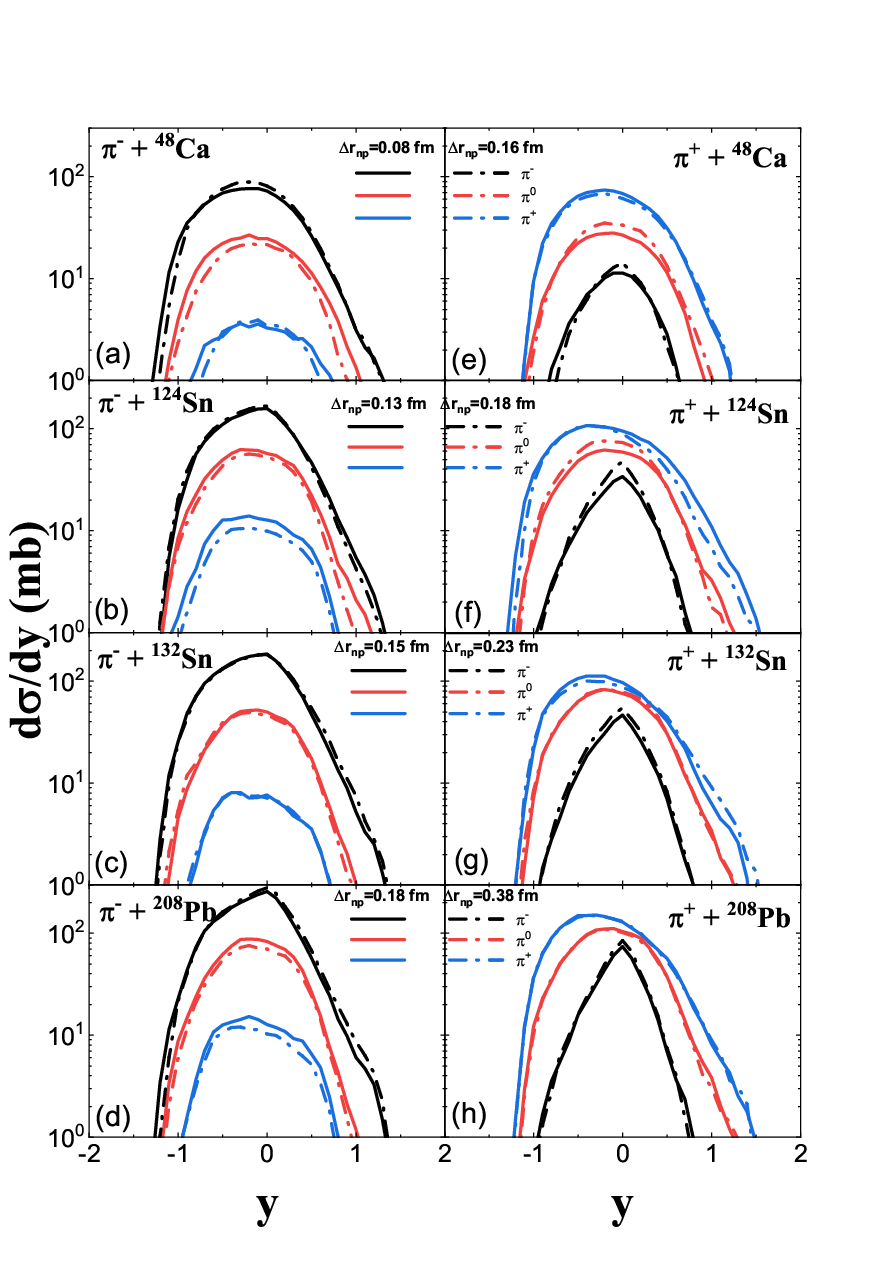}
\caption{\label{2} Rapidity distributions of $\pi^{-}$, $\pi^{0}$ and $\pi^{+}$ mesons produced in the pion-induced reactions at the incident momentum of 300 MeV/c.}
\end{figure}

\begin{figure}[ht]
\centering
\includegraphics[width=\linewidth,keepaspectratio]{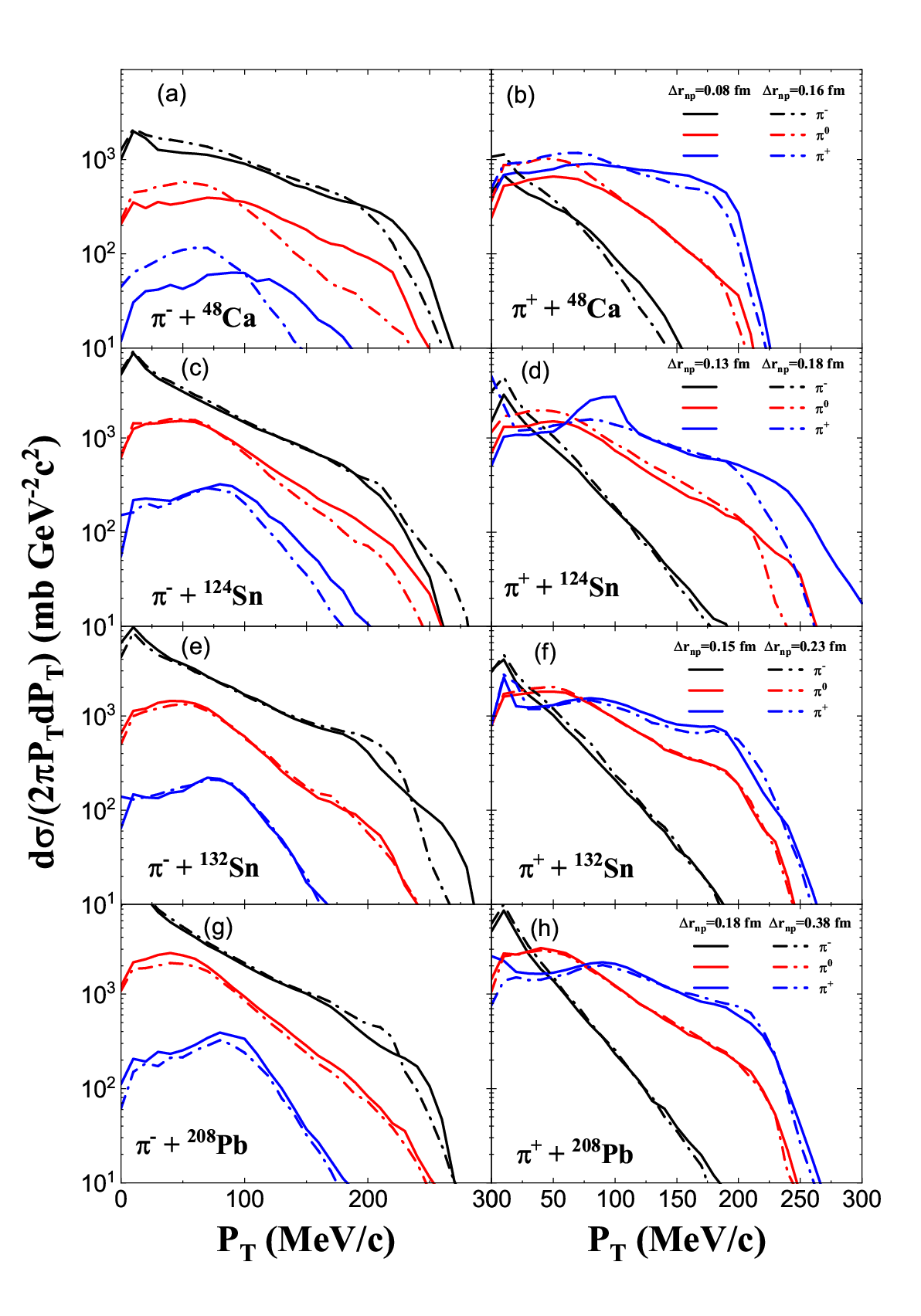}
\caption{\label{3} Transverse momentum spectra of $\pi^{-}$, $\pi^{0}$,and $\pi^{+}$ produced in the charged pion-induced reactions at 300 MeV/c. }
\end{figure}

The neutron-skin thickness plays a significant role in determining the nuclear EoS properties and is also crucial for the study of neutron-star mass-radius relation and nuclear structure, which is related to the symmetry energy at subsaturation densities. However, due to the electrically neutral nature of neutrons, it is difficulty to measure the neutron density profile inside a nucleus. Consequently, the available experimental observables for extracting the neutron-skin thickness are anticipated. In this work, we utilized the LQMD model to investigate the reactions of pion with the neutron-rich nuclei $^{48}\rm{Ca}$, $^{124,132}\rm{Sn}$ and $^{208}\rm{Pb}$. Firstly, the general observations are that both the rapidity spectra and the $\it{p_T}$ spectra depend on the neutron-skin thicknesses. Shown in Fig.\ref{2} and Fig.\ref{3} is the rapidity distribution and $\it{p_T}$ spectra of $\pi^{-}$(black), $\pi^{0}$(red) and $\pi^{+}$(blue) for $\pi^{\pm}+ ^{48}\rm{Ca}$ , $\pi^{\pm} + ^{124}\rm{Sn}$,  $\pi^{\pm} + ^{132}\rm{Sn}$,  $\pi^{\pm} + ^{208}\rm{Pb}$ collisions at an incident momentum of 300 MeV/c. The soft symmetry energy is represented by full lines, while the hard symmetry energy is denoted by the dash-dotted lines. The structure of pion production from the DCX, single-charge exchange, and elastic scattering is very similar in phase space. In Fig.\ref{2}, it can be observed that the production of $\pi^{+}$($\pi^{-}$) in the interactions of $\pi^{+}$($\pi^{-}$) with nuclei does not show a clear relationship with the neutron-skin thickness. The production of $\pi^{+}$($\pi^{-}$) is primarily associated with the elastic scattering. In the interactions of $\pi^{-}$ with nuclei that have thinner neutron skins, there is a slight increase in the production of $\pi^{+}$ and $\pi^{0}$, which is predominantly governed by DCX and single charge-exchange (SCX) processes. The DCX process $(\pi^{-}, \pi^{+})$ can be understood via the reactions associated with at least two protons, e.g., $\pi^{-} p \rightarrow \Delta^{0}$, $\quad \Delta^{0} \rightarrow \pi^{0} n$, $ \quad \pi^{0} p \rightarrow \Delta^{+}$, $\quad \Delta^{+} \rightarrow \pi^{+} n$. In nuclei with a larger neutron skin, the process is constrained because of the larger collision probabilities between $\pi^{-}$  and neutrons, which leads to the decrease of the $\pi^{+}$ and $\pi^{0}$ in the pion charge-exchange reactions. In contrast, upon considering the incidence of $\pi^{+}$ with nuclear reactions, the rapidity distributions of the pions production exhibit an inverse trend compared to those observed in $\pi^{-}$ incidence scenarios. As shown in Fig. \ref{3}, at high momentum, nuclei with thinner neutron skins exhibit larger pion production yield for both $(\pi^{-}, \pi^{0})$ and $(\pi^{-}, \pi^{+})$ reactions. Additionally, the effect of the neutron skin on the $\it{p_T}$ distribution of pions produced in charge-exchange reactions induced by $\pi^+$ is smaller compared to those induced by $\pi^-$. This indicates that in the high-momentum region, neutron-skin thickness significantly affects the reaction cross-section. In the low-momentum region, the impact of the neutron-skin thickness on the $\it{p_T}$ distribution of pions produced in charge-exchange reactions is not significant for the reactions induced by $\pi^+$ or $\pi^-$, except for the $\pi^{-} + ^{48}\rm{Ca}$.

\begin{figure*}[ht]
\centering
\includegraphics[width=\linewidth]{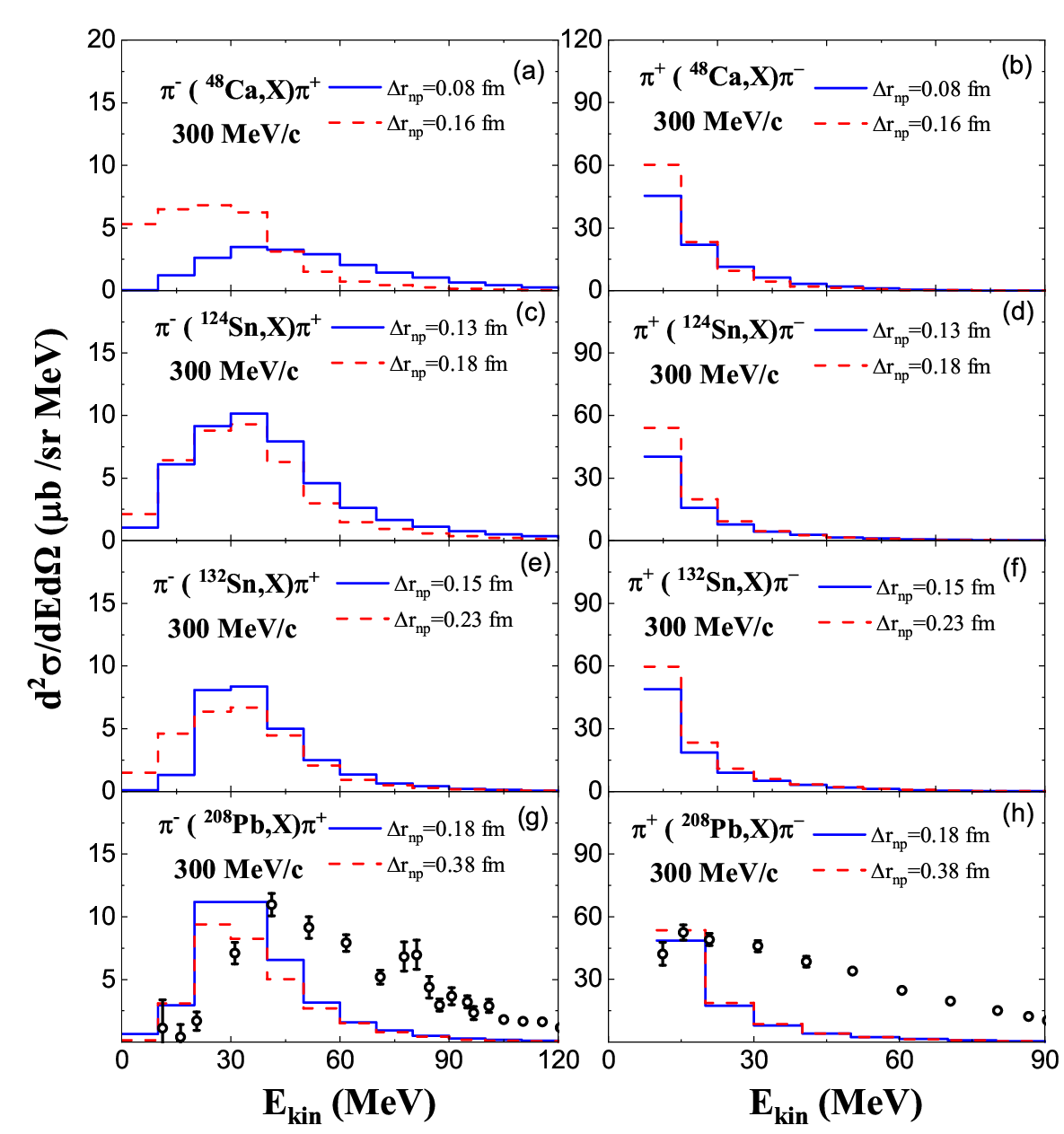}
\caption{\label{4} Kinetic energy spectra of charged pions in the double-charge exchange reactions on $^{48}\rm{Ca}$, $^{124}\rm{Sn}$, $^{132}\rm{Sn}$ and $^{208}\rm{Pb}$ nuclei at the incident momentum of 300 MeV/c, respectively. The experimental data for the reactions on $^{208}\rm{Pb}$ are shown for comparison \cite{Wo92}. }
\end{figure*}

The neutron skin have non-negligible contributions on the pions production in charge exchange reactions. Given the experimental challenges associated with measuring neutral pion ($\pi^{0}$) production via SCX reactions, our work focuses on DCX reactions, which have garnered considerable attention. In reference \cite{Bu06}, the influence of the neutron skin on the angular distribution of pions produced in the DCX process involving the reaction of $\pi^{+}$ + $^{208}\text{Pb}$ is discussed. The DCX is highly sensitive to the surface properties of nuclei because the mean-free path of incoming low-energy pions in nuclear matter is small. We employed the LQMD model to investigate the DCX reactions of $\pi^{-}$ and $\pi^{+}$ with various target nuclei ( $^{48}\rm{Ca}$, $^{124}\rm{Sn}$, $^{132}\rm{Sn}$, $^{208}\rm{Pb}$) at an incident momentum of 300 MeV/c. As shown in Fig.\ref{4}, which illustrates the kinetic energy spectra of pions produced via DCX reactions on various neutron-rich target nuclei, the effect of the neutron skin is more clearly observable. In the $(\pi^{-}, \pi^{+})$ reaction, at kinetic energies greater than 30 MeV, it is observed that the DCX cross-section is larger when $\pi^{-}$ interacts with target nuclei having a thinner neutron-skin thickness compared to the larger ones. It is attributed to the increased collision probabilities between $\pi^{-}$ and neutrons, which can affect the outcome of the DCX reactions. However, the effect of the neutron skin is observed to diminish slightly as the mass number of the target nucleus increasing. This subtle reduction may be attributed to the reabsorption of pions in nuclear medium. Pions absorption \cite{Ch89} and multiple processes \cite{Jo78} are also considered to contribute to the DCX process in pions induced nuclear reactions. In $\pi^{+}$ induced nuclear reactions, the DCX process shows an inverse behavior, at low kinetic energies, the DCX cross-section is larger when $\pi^{+}$ interacts with target nuclei that have a thicker neutron skin compared to those with a thinner neutron skin, mainly because these reactions typically occur in the neutron-rich surface regions. The energetic $\pi^{-}$ is reduced in the $\pi^{+}$ induced reaction on $^{208}\text{Pb}$ for comparison with the LAMPF data \cite{Wo92}. The difference is caused from the in-medium properties, e.g., the cross section of $\pi^{-}$-neutron scattering as shown in Eq. (14).

\begin{figure*}[ht]
\centering
\includegraphics[width=\linewidth]{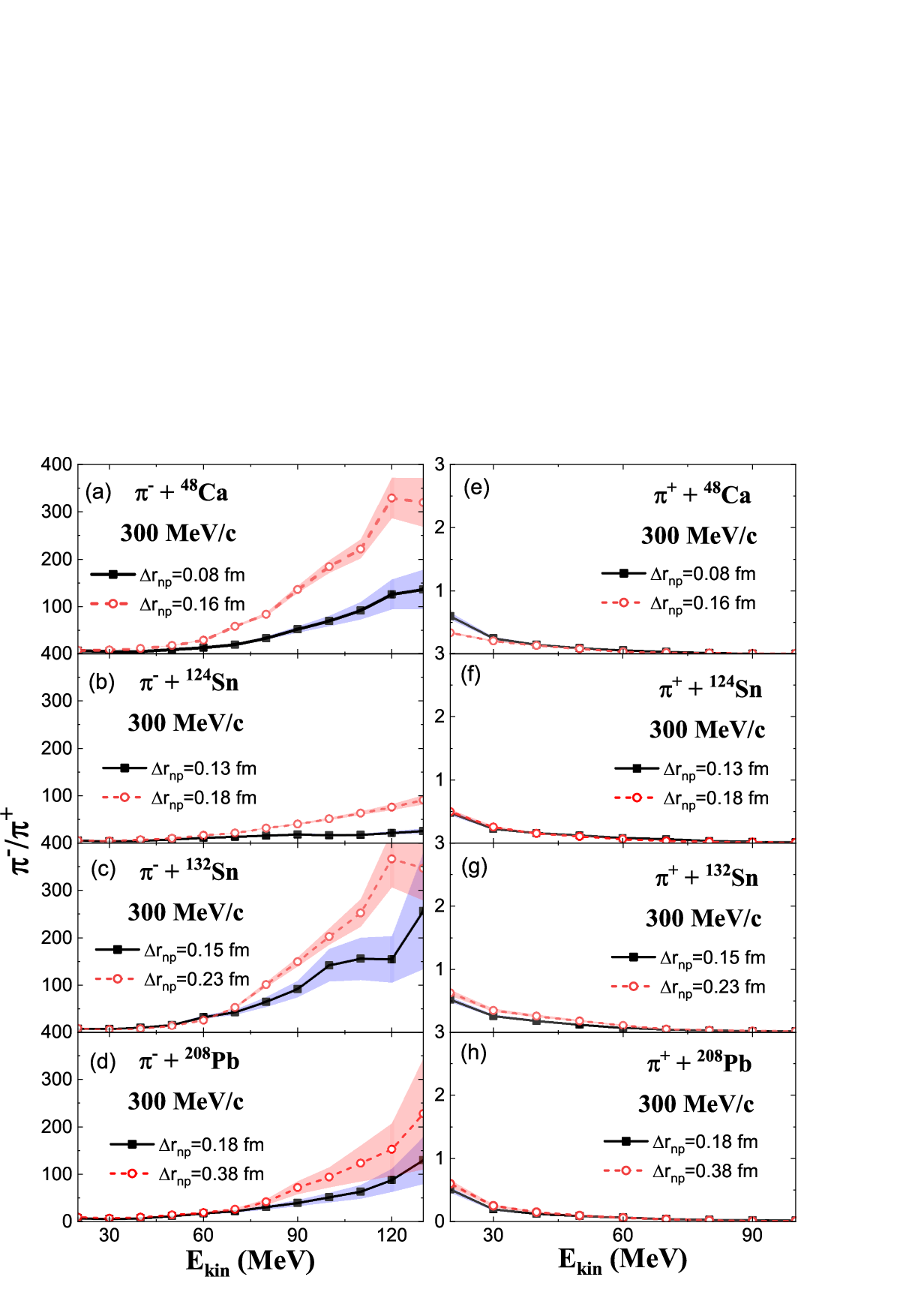}
\caption{\label{5} The influence of neutron-skin thickness on the kinetic energy spectra of $\pi^-$/$\pi^+$ ratio in the charged pion-induced reactions at 300 MeV/c. }
\end{figure*}

In the past two decades, pions produced in heavy-ion collisions have been extensively investigated for constraining the symmetry energy in the high-density region, which are associated with the in-medium properties of pions and resonances. The ratio of charged pions has been recognized as an experimental observable for the study of symmetry energy. In heavy-ion collisions, near the pion production threshold energy, the ratio of $\pi^{-}$ to $\pi^{+}$ is one of the most promising probes for symmetry energy at supra-saturated densities, as discussed in reference \cite{Ou11}, in which the impact of uncertainties in the neutron-skin thickness on extracting symmetry energy from the $\pi^{-}/\pi^{+}$ ratio was also investigated. Indeed, shortly after the advent of high-energy radioactive beams, the production of pions in peripheral nuclear reactions was proposed as a sensitive observable for extracting the neutron-skin thickness of rare isotopes \cite{Te87, Lo88, Li91}. Consequently, to further investigate the ratio of pions, we analyzed the kinetic energy spectra for the $\pi^{-}/\pi^{+}$ ratios in the pion induced reactions on different targets as shown in Fig.\ref{5}. It is obvious that for the $\pi^{-}$ induced reactions, the larger neutron-skin thickness leads to the higher $\pi^{-}/\pi^{+}$ ratio. Although the overall $\pi^{-}/\pi^{+}$ ratio appears somewhat lower in the reaction system involving $\pi^{-}$ + $^{124}\rm{Sn}$, the influence of the neutron skin on this ratio remains clearly observable. Moreover, the ratio is notably higher at high kinetic energies. For the reactions involving $\pi^{+}$ , the influence of the neutron skin on the ratio is predominantly observed in the low-energy region. The effect of neutron-skin thickness is pronounced in the $\pi^{-}$ induced reactions, in particular for the $^{48}\rm{Ca}$ based reactions.

\section{IV. Conclusions}
We have studied the DCX and  $\pi^-/\pi^+$ ratios on different targets ($^{48}\rm{Ca}$, $^{124}\rm{Sn}$, $^{132}\rm{Sn}$ and $^{208}\rm{Pb}$) at the incident momentum of 300 MeV/c within the LQMD transport model. The neutron-skin thickness is obtained by adjusting the diffuseness parameter of neutron density in the target initialization. The density dependence of symmetry energy is correlated with the neutron-skin thickness in the dynamical evolution. The transverse momentum spectra of the elastic, single-charge exchange and double-charge exchange reactions are influenced by the neutron-skin thickness. Both DCX reactions and the isospin ratio of $\pi^-/\pi^+$ are influenced by the neutron-skin thickness and symmetry energy. The larger DCX cross-section is obtained with the thinner neutron skin in the $\pi^{-}$ induced reactions. The neutron skin significantly influences the DCX process at the kinetic energies above 90 MeV. In the $\pi^{+}$ induced reactions, the DCX process manifests the opposite trend, namely, the larger DCX cross-section observed at the low kinetic energies. Overall, the $\pi^{-}/\pi^{+}$ ratio in the $\pi^{-}$ induced reactions is a nice probe for extracting the neutron-skin thickness of neutron-rich nucleus.

\textbf{Acknowledgements}
This work was supported by the National Natural Science Foundation of China (Projects No. 12175072 and No. 12311540139).

\end{document}